\documentclass[12pt,a4paper]{article}
\usepackage{amsmath,amssymb,amsfonts,amsthm,amsopn,graphicx}
\usepackage{tikz} 
\usetikzlibrary{arrows}

\usepackage{xcolor}

\usepackage{showlabels}

\allowdisplaybreaks[3]
\numberwithin{figure}{section}

\newcommand{\ds}{\displaystyle}

\renewcommand{\author}[1]{\large\rm #1\\ \bigskip}
\newcommand{\address}[1]{{\normalsize\it #1\\}\bigskip}
\renewcommand{\title}[1]{\bigskip\bigskip\Large\bf #1\bigskip\bigskip\\}

\usepackage[body={15.5cm,22cm}]{geometry}
\begin{document}
\vglue 2cm

\begin{center}

\title{Bethe Ansatz and Rogers-Ramanujan-type identities.}
\author{Sergey M.~Sergeev.}

\vspace{.5cm}

\address{Department of Theoretical Physics,
         Research School of Physics and Engineering,\\
    Australian National University, Canberra, ACT 0200, Australia\\
    and\\
   Faculty of Science and Technology, \\
   University of Canberra, Bruce ACT 2617, Australia }


\end{center}
\begin{abstract}
The Rogers-Ramanujan identity for $\!\!\phantom{|}_1\psi_1$
$$
\sum_{n\in\mathbb{Z}} \frac{(a;q)_n}{(b;q)_n} z^n\;=\;
\frac{(q,b/a,az,q/az;q)_\infty}{(b,q/a,z,b/az;q)_\infty}
$$
can be classified as one related to the Bethe Ansatz for ``chain length $N=1$, ground state XXZ model with an arbitrary negative spin''.
\end{abstract}



In this note we observe some exotic way to produce the spectrum of auxiliary transfer matrices for $\mathcal{U}_q(\widehat{sl}_2)$ models. This way is equivalent to the standard Bethe Ansatz, however it uses different variables. As a by-product, the method is related to generalisations of the Rogers-Ramanujan identity \cite{GR} mentioned in the  Abstract. Note, we observe the Rogers-Ramanujan identities of a type different from those obtained by Andrews, Baxter and Forrester \cite{ABF}.

Our staring point is the Baxter's equation,
\begin{equation}\label{TQ1}
t(x) Q(x) \;=\; (1-\xi x)^N Q(qx) \:+\; \omega q^S (\xi-x)^N Q(x/q)\;,
\end{equation}
where $t(x)$ stands for the auxiliary transfer-matrix,
\begin{equation}\label{t}
t(x)\;=\; 1+\omega q^S\xi^N + t_1 x+ \cdots + t_{N-1} x^{N-1} + \; (\omega+q^S\xi^N) (-x)^N\;,
\end{equation}
$N$ is the chain length, $S$ is ``total spin'', $\omega$ is a field, and $Q(x)$ is a polynomial of degree $S$. Standard finite-dimensional case corresponds to $\xi=q^{-j}$ where $j$ is a half-integer local spin. However, we assume here another regime,
\begin{equation}
\ds |q|\;<\;1\;,\quad |\xi|\;<\;1\;.
\end{equation}
Lowest and highest terms of $t(x)$ as well the other terms in (\ref{TQ1}) come from unessential details of properly defined $\mathcal{U}_q(\widehat{sl}_2)$ $L$-operators.
The Bethe-Ansatz equations 
\begin{equation}\label{BAE}
(1-\xi x)^N Q(qx) \:+\; \omega q^S (\xi-x)^N Q(x/q)\;=\;0\quad \textrm{on}\quad Q(x)\;=\;0\;,
\end{equation}
are regarded as a closed system of equations for $S$ zeros of $Q(x)$, solution of (\ref{BAE}) provides eigenvalue of $t(x)$.

However, equation (\ref{TQ1}) allows one to introduce different than $Q(x)$ objects. Consider two more $TQ$-type equations,
\begin{equation}\label{TH}
\left\{
\begin{array}{l}
\ds t(x) H(x) \;=\; H(x/q) \;+\; \omega q^S (1-\xi x)^N (\xi-qx)^N H(qx)\;,\\
\\
\ds t'(x) H'(x) \;=\; H'(qx) \;+\; \omega^{-1} q^S (1-\xi/x)^N (\xi-q/x)^N H'(x/q)\;,
\end{array}\right.
\end{equation}
where
\begin{equation}
t'(x) \;=\; t(x)/\omega (-x)^{N}\;,
\end{equation}
and
\begin{equation}
H(x)\;=\;1+h_1^{}x + h_2^{}x^2+\cdots\;,\quad 
H'(x)\;=\;1+h_1'/x+h_2'/x^2+\cdots\;.
\end{equation}
For an arbitrary $t(x)$, $H$ and $H'$ are the uniquely defined seria convergent everywhere except $x=\infty$ and $x=0$ correspondingly.
Moreover, the functions $H$ and $H'$ have the remarkable matrix product representations. Let
\begin{equation}
L(x)\;=\;\left(\begin{array}{cc}
\ds t(x) & -\gamma(x) \\
\ds 1 & \ds 0 \end{array}\right)\;,\quad
L'(x) \;=\; \left(\begin{array}{cc}
\ds t'(x) & \ds 1 \\
\ds -\gamma'(x) & \ds 0
\end{array}\right)\;,
\end{equation}
where (see (\ref{TH}))
\begin{equation}
\gamma(x)\;=\;\omega q^S (1-\xi x)^N (\xi-qx)^N\;,\quad
\gamma'(x)\;=\;\omega^{-1} q^S (1-\xi/x)^N (\xi-q/x)^N\;.
\end{equation}
The following semi-infinite matrix products are well defined:
\begin{equation}
L(x) L(qx) (q^2x) \cdots \;=\; \frac{1}{1-\omega q^S\xi^N} \left(\begin{array}{c}
\ds H(x/q)\\ H(x) \end{array}\right) \left(1, -\omega q^S\xi^N \right)\;,
\end{equation}
and
\begin{equation}
\cdots L'(x/q^2) L'(x/q) L'(x) \;=\; \frac{1}{1-\omega^{-1}q^S\xi^N} \left(\begin{array}{c}
\ds 1 \\ \ds - \omega^{-1} q^S\xi^N\end{array}\right) \left(H'(qx), H'(x) \right)\;.
\end{equation}
Next, for general $t(x)$, the Wronskian of $H,H'$,
\begin{equation}
\Theta(x)\;=\;H'(x) H (x/q) - q^S (\xi-x)^N (\xi-q/x) ^N H'(x/q) H(x)\;,
\end{equation}
satisfies
\begin{equation}
\Theta(x) \;=\; \omega (-x)^N \Theta(qx)\;,
\end{equation}
so that the $\theta$-function expansion holds:
\begin{equation}
\Theta(x) \;=\; \Theta_0\; \prod_{k=1}^N (x/z_k;q)_\infty (qz_k/x;q)_\infty\;,\quad \prod_{k=1}z_k\;=\;\omega^{-1}\;.
\end{equation}
Matrix product representation allows one to write $\Theta(x)$ as a matrix product infinite in both directions.
The set of $\{z_k\}$ is the set of variables alternative to the standard Bethe variables, i.e. to the set of zeros of $Q(x)$. The real aim of this note is to reformulate the Bethe Ansatz equations (\ref{BAE}) in the terms of $\{z_k\}$.

To do this, let us assume (\ref{TQ1}) and compute the Wronskians of $H^{\#}$ and $Q$:
\begin{equation}\label{HQ}
\left\{
\begin{array}{l}
\ds H(x/q) Q(x) - \omega q^S (\xi-x)^N H(x) Q(x/q) \;=\; (1-\omega q^S\xi^N) (\xi x;q)_\infty^N\;,\\
\\
\ds H'(x) Q'(x/q) - \omega^{-1} q^S (\xi-q/x)^N H'(x/q) Q'(x) \;=\; \kappa (1-\omega^{-1}q^S\xi^N) ( q\xi/x;q)_\infty^N\;,
\end{array}\right.
\end{equation}
where $Q(0)=1$, $Q'(x) \;=\; Q(x)/x^S$ and $\kappa=Q'(\infty)$. Consequence of (\ref{HQ}) is
\begin{equation}\label{Q2}
Q(x)\;=\;(1-\omega q^S\xi^N)\frac{H'(x) (\xi x;q)_\infty^N}{\Theta(x)} + \kappa x^S 
(1-\omega^{-1} q^S \xi^N) \frac{H(x) (\xi/x;q)_\infty^N}{\Theta(qx)}\;.
\end{equation}
Thus we obtain the quantisation condition equivalent to the initial Bethe-Ansatz equations (\ref{BAE}). Namely, for given arbitrary $t(x)$, one computes $H^{\#}(x)$ and therefore $\{z_k\}$. Then, the spectral equations, alternative but equivalent to (\ref{BAE}), are
\begin{equation}\label{BAE2}
H'(x) (\xi x;q)_\infty^N + \kappa' x^S (-x)^N H(x) (\xi/x;q)_\infty^N\;=\;0\quad \textrm{on}\quad 
x\in \{z_k\}\;,
\end{equation}
where $\kappa'$ in this approach is an extra parameter, i.e. in fact the number of equations in (\ref{BAE2}) is $N-1$ in accordance to the number of independent $\{z_k\}$, $\prod z_k = \omega^{-1}$.

Note that for the case of the modular double and quantum dilogarithms, when polynomial $Q(x)$ is undefined, equations of the type (\ref{BAE2}) play the role of the major spectral equations, see \cite{S1,S2,S3,F}.

Now suppose that all spectral equations are solved, i.e. (\ref{BAE}) or (\ref{BAE2}) are satisfied, and equation (\ref{Q2}) becomes an identity. Equations (\ref{HQ}) can be solved with respect to $H^{\#}$,
\begin{equation}
H(x/q) \;=\; (1-\omega q^S\xi^N) Q(x/q) (\xi x;q)_\infty^N \sum_{n=0}^\infty \frac{\ds (\omega q^S\xi^N)^n}{Q(q^{n-1}x) Q(q^nx)} \frac{(x/\xi;q)_n^N}{(\xi x;q)_n^N}\;,
\end{equation}
and
\begin{equation}
H'(qx) \;=\; \kappa(1-\omega^{-1}q^S\xi^N) Q'(qx) (\xi/x;q)_\infty^N \sum_{n=0}^\infty \frac{\ds (\omega^{-1}q^S\xi^N)^n}{Q'(q^{n+1}x) Q'(q^nx)} \frac{(1/\xi x;q)_n^N}{(\xi/x;q)_n^N}\;.
\end{equation}
Using this, one can bring the identity (\ref{Q2}) to the form
\begin{equation}\label{RR}
\sum_{n\in \mathbb{Z}} \frac{\ds (\omega q^S\xi^N)^n}{Q(q^{n-1}x) Q(q^nx)} 
\frac{(x/\xi;q)_n^N}{(\xi x;q)_n^N}\;=\;
\frac{(q/x)^S}{\kappa (1-\omega q^S\xi^N) (1-\omega^{-1} q^S \xi^N)} \; \frac{\ds \Theta(x)}{\ds (\xi x;q)_\infty^N (q\xi/x;q)_\infty^N}\;.
\end{equation}
In particular, in the case $N=1$ and $S=0$, $Q(x)=1$, equation (\ref{RR}) becomes the Rogers-Ramanujan identity mentioned in the Appendix (we did not discuss the normalisation constant $\Theta_0$).

Relation (\ref{RR}) is a particular case of more general one,
\begin{equation}\label{RRgen}
\psi(x)\;\stackrel{def}{=}\;
\sum_{n} z^n f(q^n x) \prod_{k=1}^N \frac{\ds (x/a_k;q)_n}{\ds (x/b_k;q)_n}\;=\;
\frac{\ds W(x)}{\ds \prod_{k=1}^N (x/b_k;q)_\infty (qa_k/x;q)_\infty}\;,
\end{equation}
where, since
\begin{equation}
\psi(x) \;=\; z \prod_{k=1}^N \frac{(1-x/a_k)}{(1-x/b_k)} \; \psi(qx)\;,
\end{equation}
function $W(x)$ satisfies
\begin{equation}
W(x)\;=\; z\frac{(-x)^N}{\prod_k a_k} \; W(qx)\;.
\end{equation}
This note gives an example of way how to avoid extra poles of $f(x)$ in (\ref{RRgen}), what is the Bethe Ansatz equations (\ref{BAE}), so that $W(x)$ becomes a theta-function, and clarifies the structure of this theta-function.

\bigskip

\noindent
\textbf{Acknowledgement.} I would like to thank Vladimir Bazhanov and Vladimir Mangazeev for valuable discussions.
Also I acknowledge the support of the Australian Research Council grant 
DP190103144.

\end{document}